\documentclass[a4paper,12pt]{article}
\usepackage[russian]{babel}
\usepackage{amsmath}
\usepackage{ulem}
\usepackage{calc}
\usepackage{vmargin}
\usepackage{amstext}
\usepackage{graphicx}
\usepackage{amsfonts}
\usepackage{amssymb}
\usepackage{titlesec}

\begin{document}


\title{ Поверхностные гравитационные волны уравнения Буссинеска }

\author{О.В. Капцов\\ Институт вычислительного моделирования, Россия \\ email: kaptsov@icm.krasn.ru \\ \\
	Д.О. Капцов\\ Институт вычислительного моделирования, Россия \\ email: hot.dok@gmail.com }
\date{}
\maketitle

\begin{abstract}
	
В работе рассматривается классическое уравнение Буссинеска, описывающее гравитационные волны на мелкой воде. Для построения точных решений используется билинейное представление Хироты. Найденные решения описывают, в частности, волновые пакеты, волны на солитонах, "танцующие волны". Указывается принцип умножения решений уравнения Хироты, позволяющие строить более сложные структуры из солитонов, волновых пакетов и других типов волн.

\end{abstract}

\section{Введение}

Семидесятые и восьмидесятые годы прошлого века были периодом бурного развития солитонной тематики, затронувшей различные области нелинейной физики. Были разработаны новые и возрождены забытые методы интегрирования ряда нелинейных уравнений \cite{Абловиц}, \cite{Захаров}, \cite{Матвеев}. К сожалению, число уравнений, интегрируемых этими методами, не слишком велика. Следует отметить, что сам термин "интегрируемость" \ имеет различные толкования, так как  единого устоявшегося определения не существует. 

Наибольший интерес среди интегрируемых уравнений вызывают модели, обладающие солитонными решениями. К ним относится известное уравнение Буссинеска \cite{Уизем,Debnath}
\begin{equation} \label{BusOrig}
\eta_{tt} = gh_0 \eta_{xx} + \frac{3}{2}g(\eta^2)_{xx} +
\frac{1}{3}gh_0^3\eta_{xxxx},
\end{equation}
где $g$ -- ускорение свободного падения; $h_0$ -- глубина невозмущенной жидкости;
$\eta$ -- функция от $t$ и $x$, описывающая отклонение поверхности воды от невозмущенного состояния. Солитонные решения этого уравнения найдены Хиротой \cite{Hirota}, бризерные решения указаны в \cite{Tajiri}, а рациональные в \cite{Абловиц,Clarkson}. Бризеры получаются из солитонов комплисификацией параметров, при этом сами решения должны оставаться вещественными. Сингулярные решения уравнения Буссинеска рассматривались в \cite{Bogdanov}.
В работе \cite{Капцов} было замечено, что выбирая подходящим образом комплексные волновые параметры, можно получить различные типы волн уравнения Буссинеска. 

В данной работе развиваются исследования, начатые в \cite{Капцов}. Во втором параграфе с помощью линейных дифференциальных связей четвертого порядка находится решение равнения Буссинеска, зависящее от нескольких произвольных констант и выражающееся через элементарные функции. Из бризерных решений выделяются специальные типы волн, такие как волновые пакеты, "танцующие" волны и волны на солитонах. Представляется удобная форма принципа суперпозиции, позволяющая получать из специальных типов волн более сложные структуры. В частности, среди них имеются структуры, описывающие взаимодействие волновых пакетов с "танцующими"\ солитонами и между собой.
В третьем параграфе найдены новые решения уравнения Буссинеска, похожие на решения из параграфа 2. Однако эти решения удовлетворяют другим дисперсионным соотношениям и стремятся к другой константе при $|x|\rightarrow \infty$.

\section{Базовые решения и их суперпозиция}

Уравнение Буссинеска (\ref{BusOrig}) приводится к виду

\begin{equation} \label{Bus}
w_{tt} = w_{xx} + 3(w^2)_{xx} + w_{xxxx}
\end{equation}
преобразованием растяжения. Введем новую функцию $u$, полагая $w = u_{xx}$. Тогда уравнение (\ref{Bus}) записывается в виде:
$$ \frac{\partial ^2}{\partial x^2}(u_{tt} - u_{xx} - 3u_{xx}^2 - u_{xxxx}) = 0 .
$$

В дальнейшем считаем, что функция $u$ удовлетворяет уравнению 
$$ u_{tt} = u_{xx} + 3u_{xx}^2 + u_{xxxx} . $$
Замена $ u = 2\ln(H) $ приводит последнее уравнение к билинейному представлению 
\begin{equation} \label{BusBilin}
HH_{tt} - H_{t}^2 - HH_{xxxx} + 4H_xH_{xxx} - 3H_{xx}^2 - HH_{xx} + H_x^2 = 0.
\end{equation}
Таким образом, решения уравнения (\ref{Bus}) получаются из решений уравнения (\ref{BusBilin}) по формуле
\begin{equation} \label{Transform}
w=2\frac{\partial ^2}{\partial x^2}\ln H .
\end{equation}
Эта формула предложена Хиротой \cite{Hirota}, он же использовал билинейное уравнение, равносильное (\ref{BusBilin}), для построения многосолитонных решений уравнения Буссинеска. В частности, тройка решений уравнения (\ref{BusBilin}) имеет вид
\begin{equation} \label{Soliton123}
\begin{split}
H_1 = & 1 + f_1, \\
H_2 = & 1 + f_1 + f_2 + p_{12}f_1f_2, \\
H_3 = & 1 + f_1 + f_2 + f_3 + p_{12}f_1f_2 + p_{13}f_1f_3+ p_{23}f_2f_3 + p_{12}p_{13}p_{23}f_1f_2f_3,
\end{split}
\end{equation}
где $ f_i = s_i \exp(k_ix + m_it),$ где $ s_i, k_i $ -- произвольные константы, а 
\begin{equation} \label{p}
p_{ij} = \frac{ 3(k_i - k_j)^2 + (m_i - m_j)^2}{3(k_i + k_j)^2 + (m_i - m_j)^2 }.
\end{equation}
При этом параметры $ k_i, m_i $ удовлетворяют "дисперсионному" соотношению
\begin{equation} \label{m}
m_i^2 = k_i^2 + k_i^4 .
\end{equation}
Функции $H_1, H_{2} $ являются решениями уравнения четвертого порядка 
\begin{equation} \label{h2solution}
\frac{\partial }{\partial x}(\frac{\partial }{\partial x} - k_1)(\frac{\partial }{\partial x} - k_2)(\frac{\partial }{\partial x} - k_1 - k_2)H = 0.
\end{equation}

Возникает вопрос: имеются ли отличные от $H_1, H_2 $ решения системы (\ref{Bus}), (\ref{h2solution})? Пусть $ k_1 = k_2 = k \neq 0 \in R. $ Тогда общее решение уравнения (\ref{h2solution}) имеет вид

\begin{equation} \label{h2common}
H = c_1 + (c_2x + c_3)e^{kx} + c_4e^{2kx},
\end{equation}
где $ c_1,\dots,c_4 $ -- некторые функции от $t$. Не ограничивая общности, $ c_1 $ можно считать равной 1 или 0. Предположим, что 
$ c_1 = 1 $. Подставляем представление (\ref{h2common}) в уравнение 
(\ref{Bus}) и получаем переопределенную систему обыкновенных дифференциальных уравнений относительно функций $ c_2, c_3, c_4 $. Решая эту систему, находим 
$$
c_2 = -\frac{sme^{-mt} }{k(2k^2+1)}, \qquad c_3 = se^{-mt},\qquad c_4 = - \frac{ s^2(4k^2+3)e^{-2mt} }{ 12k^2(2k^2+1)^2 }, 
$$
где $ s \in R, m^2 = k^2 +k^4 $. Другие решения системы (\ref{Bus}), (\ref{h2solution}) представлены в \cite{Капцов}.

Обобщением (\ref{h2solution}) является дифференциальное уравнение порядка $ 2^n $:
$$
\frac{\partial }{\partial x} \prod_{\substack{1 \le p \le n \\ 1 \le i_1 < ... < i_p \le n }} (\frac{\partial }{\partial x} - k_{i_1} - ... - k_{i_p} )H = 0  .
$$
Решая это уравнение при разнообразных параметрах $ k_1, ..., k_n $ и подставляя в (\ref{Bus}), можно найти солитонные, рациональные, бризерные и другие решения уравнения Буссинеска. 

Далее мы сосредоточим свое внимание на специальном классе решений. Этот класс заслуживает особого изучения, так как он задает интересные типы волн. 
\textbf{Определение. } 
Функцией Хироты (или $H$-функцией) будем называть функцию вида

\begin{equation} \label{Hfunc}
H(k_1, ... , k_n) = ((1 + f(k_1))*...*(1 + f(k_n))),
\end{equation}
где $ k_i $ -- комплексные параметры, $ f(k_i) = \exp(k_ix + \sqrt{k_i^2 + k_i^4}t)$. Символ * означает бинарную операцию, удовлетворяющую условиям коммутативности, ассоциативности, дистрибутивности и заданную условиями $ \forall c \in\mathbb{C} $
$$
c * f(k_i) = cf(k_i), \qquad f(k_{i_1}) * ... * f(k_{i_l}) = p_{i_1 ... i_l}  f(k_{i_1})...f(k_{i_l}),
$$
где $ p_{i_1...i_l} = p_{i_1i_2} \cdot \cdot \cdot p_{i_{l-1} i_l}  $ -- произведение всех $ p_{i_ri_q}$ $ (r < q \le l) $ заданных формулой (\ref{p}).

Сам Хирота использовал для многосолитонных решений другое представление \cite{Hirota}. 

Очевидно, функция Хироты $ H(k_1, ... , k_n) $ не меняется при перестановке любых двух параметров $ k_i, k_j $. Если же какая то пара параметров совпадает, то функция Хироты тождественно равна нулю.

Легко видеть, что функции Хироты образуют бесконечную коммутативную полугруппу с умножением

\begin{equation} \label{Hgroup}
H(k_1, ... , k_n) * H(k_1', ... , k_m') = H(k_1, ... , k_n, k_1', ... , k_m') .
\end{equation}

Из этого следует принцип суперпозиции: если заданы две функции Хироты, то их * произведение также является функцией Хироты  и решением уравнения (\ref{BusBilin}).

\begin{figure}[h!]
	\centering
	\includegraphics[height=5cm, width=10cm]{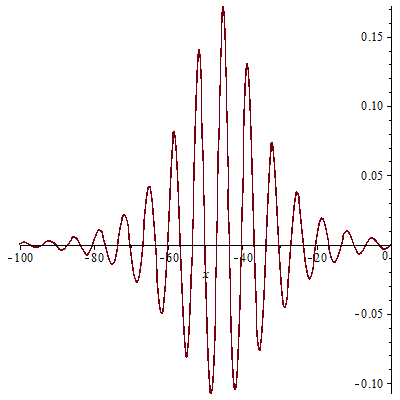}
	\caption{Волновой пакет}
	\label{fig:WavePacket}
\end{figure}

\begin{figure}[h!]
	\centering	
	\includegraphics[height=5cm, width=10cm]{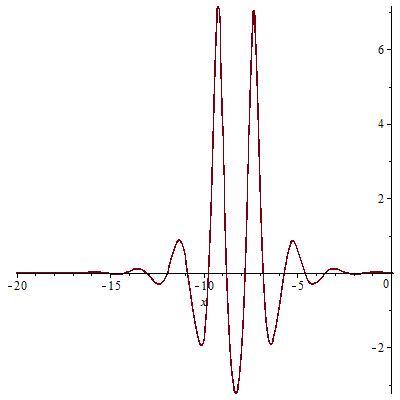}
	\caption{Танцующие волны}
	\label{fig:DancingWave}
\end{figure}

\begin{figure}[h!]
	\centering	
	\includegraphics[height=5cm, width=10cm]{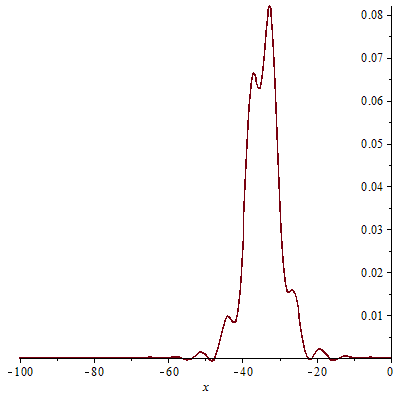}
	\caption{Волны на солитоне}
	\label{fig:SolitonWaves}
\end{figure}

Нас будет интересовать только вещественные значения функции Хироты, которые могут зависеть от комплексных параметров. Начнем со случая двух параметров $ k_1 = a + ib, k_2 = a - ib$. Очевидно, комплексно сопряженные параметры задают вещественную функцию $ H(k_1, k_2)$. Мы выделяем три типа функций $ H $, задающих три типа  волн для  уравнения Буссинеска (\ref{Bus}). К первому типу волн относятся волновые пакеты \ (рис. \ref{fig:WavePacket}), ко второму (D-волны) -- "танцующие волны" \ (рис. \ref{fig:DancingWave}), к третьему 
(R-волны) -- волны на солитоне  (рис. \ref{fig:SolitonWaves}).


Константы $ a, b$, задающие комплексно сопряженные числа $ k_1, k_2 $, таковы
\begin{center}
	\begin{tabular}{ccc}
		a = 0.1 , & b = 0.933 & для P-волны, \\
		a = 0.2 ,& b = 0.933 & для D-волны, \\
		a = 0.911, & b = 2.8213 & для R-волны. \\
	\end{tabular}
\end{center}


В каждый фиксированный момент времени $ t $ решения уравнения Буссинеска (\ref{Bus}), отвечающие $ P, D, R $-волнам стремятся к нулю при $ |x| \rightarrow \infty$. 
 Анимационные картинки этих волн представлены в P.gif, D.gif, R.gif. Кроме указанных выше значений констант $ a, b $, существуют области на плоскости $ R^2(a,b)$, для которых реализуются $P,R,D$ - волны. Заметим, что обычному солитону (или S-волне) соответствует функция Хироты $ H(k) $ с вещественным параметром $ k \neq 0$.

Для построения более сложных волновых структур из $P, D, R, S $-волн достаточно использовать формулу умножения (\ref{Hgroup}) $H$ функций, соответствующих этим волнам и (\ref{Transform}). На SD.gif приведены анимационные картины, показывающие взаимодействие солитона (параметр $k = 1$) с $D$ - волной (параметры указаны выше). Как видно на картинках форма и скорость волн восстанавливаются после взаимодействия. Следует выделить интересную структуру (SP2.gif), состоящую из волнового пакета и солитона (параметры  $ k_1=0.1 + 0.933i,\ k_2 = \overline{k_1},\ k_3=1.5 $). Солитон и волновой пакет движутся друг за другом, фактически не взаимодействуя. На PP.gif показаны взаимодействия двух волновых пакетов между собой. Волновые числа таковы: $k_1 = 0.1 + 0.933i,\ k_2 = \overline{k_1}, k_3 = 0.1 + 1.933i,\ k_4 = \overline{k_3} $. На PP2.gif показано движение двух фактически не взаимодействующих волновых пакетов.
Таким же образом можно получить решение уравнения Буссинеска, описывающее движение любого числа $ P, D, R, S $-волн.

\section{Новые дополнительные решения}

Множество решений уравнения Буссинеска выражающихся через элементарные функции, далеко не исчерпывается приведенными выше. В этом параграфе мы построим новые решения, вновь используя билинейное уравнение.

Выполним замену $ w = v- \frac{1}{6} $ и приведем уравнение (\ref{Bus}) к виду
\begin{equation} \label{bus_subs}
v_{tt} = (3v^2 + v_{xx})_{xx}.
\end{equation}

Затем введем новую функцию $z$ по формуле $ v = z_{xx}$ и получим уравнение шестого порядка, которое два раза интегрируется. В результате приходим к уравнению 
$$
z_{tt} = 3z_{xx}^2 + z_{xxxx}
$$
После замены $z=2\ln(G)$ получаем новое билинейное уравнение 
\begin{equation} \label{bus_new_line}
GG_{tt} - G_{t}^2 - GG_{xxxx} + 4G_{x}G_{xxx} - 3G_{xx}^2 = 0,
\end{equation}
отличающееся от (\ref{BusBilin}) только двумя слагаемыми. 

Введем функции $g_i = \exp(k_ix + k_i^2t),\ k_i \in \mathbb{C}$, $i=1,2,3.$ 
Прямыми вычислениями несложно проверить, что функции 
$$G_1 = 1+g_1,\qquad G_2 = 1+g_1+g_2+p_{12}g_1g_2,$$
$$G_3 = 1 + g_1 + g_2 + g_3 + p_{12}g_1g_2 + p_{13}g_1g_3 + p_{23}g_2g_3 + p_{12} p_{13} p_{23}g_1g_2g_3$$
удовлетворяют билинейному уравнению (\ref{bus_new_line}), если 
$$
p_{ij} = \frac{(k_i-k_j)^2}{k_i^2+k_ik_j+k_j^2}, \qquad (i \leq i < j \leq 3).
$$
Значит, функции 
$$
w_i = -\frac{1}{6} + 2\frac{\partial^2}{\partial x^2}\ln(G_i), \qquad i=1,2,3,
$$
являются решениями уравнения  Буссинеска.

\begin{figure}[h!]
	\centering
	\includegraphics[height=5cm, width=10cm]{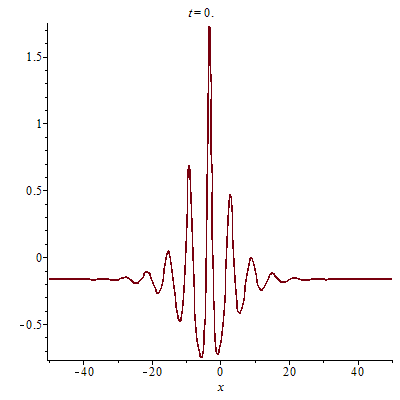}
	\caption{Волновой пакет}
	\label{fig:PWnew}
\end{figure}

Несмотря на то, что эти решения $w_i$ отличаются от решений, приведенных во втором параграфе, качественное поведение тех и других похожее.
Если параметры $k_1, k_2, k_3$ вещественных попарно различны и не равны нулю, то соответствующие функции $ w_i$ $(i=1,2,3)$ задают
одно-, двух- и трехсолитонные решения уравнения (\ref{Bus}) такие, что $w_i \rightarrow -\frac{1}{6}$ при $|x|\rightarrow \infty $. 
Если взять параметры $k_1, k_2$ комплексносопряженными, то решение $w_2$ будет вещественным. Например, при $k_1=0,2+i, k_2=0,2-i$
мы получаем волновой пакет, график которого представлен на рис. \ref{fig:PWnew} при $t=0$. При $k_1=0.5+i,\ k_2=0.5-i$, приходим к решению, подобному $D$-волне.

Отметим, что решения уравнения Буссинеска (\ref{Bus}) можно искать в виде:
$$
w=A + 2\frac{\partial^2}{\partial x^2}\ln K, 
$$
где $A$ -- произвольная константа, а функция $K(t,x)$ удовлетворяет билинейному уравнению
$$
KK_{tt}-K_t^2-KK_{xxxx}+4k_xK_{xxx}-3K_{xx}^2-AKK_{xx}+AK_x^2=0.
$$
Следуя методике описанной выше, несложно построить новые солитоноподобные решения уравнения Буссинеска (\ref{Bus}).

Интересно было бы рассмотреть другой вариант уравнения Буссинеска
$$
w_{tt} = w_{xx} + 3(w^2)_{xx}+w_{ttxx}
$$
с нулевыми граничными условиями при $ |x| \mapsto \infty$ и начальными условиями в виде $ P, D$ или $R$-волн. Возникает вопрос об эволюции решений данной задачи и о том, будут ли они близки к соответствующим решениям уравнения (\ref{Bus}) в течении долго времени.

Работа выполнена при финансовой поддержке РФФИ (грант 17-01-00332-а).

\end{document}